\documentclass{llncs}

\usepackage{amssymb}
\usepackage{amsmath}
\usepackage{color}
\usepackage{graphicx}
\usepackage{graphics}

\newtheorem{rem}{Remark}

\newcommand{\RR}{{\mathbb R}}

\newcommand{\norm}[1]{\lVert#1\rVert}

\newcommand{\dotex}{{\frac{d}{dt}_{t=0}}}

\def \ee {\begin{equation}}
\def \eee {\end{equation}}
\def \ea {\begin{align}}
\def \eea {\end{align}}
\def \eqe {\begin{eqnarray}}
\def \eqee {\end{eqnarray}}

\begin{document}

\title{An intrinsic Cram\'er-Rao bound on Lie groups}
\author{Silv\`{e}re Bonnabel and Axel Barrau}
\institute{MINES ParisTech, PSL Research University, Centre for robotics, 60 Bd St Michel 75006 Paris, France
  \email{[axel.barrau,silvere.bonnabel]@mines-paristech.fr} }

\maketitle

\begin{abstract}
In his 2005 paper, S.T. Smith proposed an intrinsic Cram\'er-Rao bound on the variance of estimators of a parameter defined on a Riemannian manifold. In the present technical note, we consider the special case where the parameter lives in a Lie group.  In this case, by choosing ,e.g., the right invariant metric, parallel transport becomes very simple, which allows a more straightforward and natural derivation of the bound in terms of Lie bracket, albeit for a slightly different definition of the estimation error. For bi-invariant metrics, the Lie group exponential map we use to define the estimation error, and the Riemannian exponential map used by S.T. Smith coincide, and we prove in this case that both results are identical indeed. 
\end{abstract}

\section{Introduction}

The Cram\'er-Rao bound is a lower bound on the achievable precision of any {unbiased} estimator of a vector $\theta$ which parametrizes a family of probability distributions $p(X|\theta)$, from a sample $X_1,\cdots,X_n$. This bound is standard in classical estimation theory. Differential geometry considerations in statistics can be traced back to  equivariant estimation  \cite{pitman1939estimation}  (see also \cite{berger1985statistical} and references therein for a more recent exposure) and of course to the work by Fisher on the Information metric, and all the works that followed, notably in information geometry.  
The paper \cite{smith-2005} proposes to derive an intrinsic Cram\'er-Rao bound for the case where the parameter lives in a Riemannian manifold. The two examples given are subspace estimation (that pertains to the Grassman manifold) and covariance matrix estimation (that pertains to the cone of positive definite matrices), both examples being related to signal processing applications. See also the nice extensions proposed since then by N. Boumal \cite{boumal2013intrinsic,boumal2014cramer}, and our gentle introduction to the subject \cite{barrau2013note} for more details. 

Our motivating example is the so-called  Wahba's problem \cite{wahba1965}, named after  Grace Wahba, which is an optimization problem where the  parameter is  a rotation matrix, but which can also be viewed as the search for a maximum likelihood rotation estimator. The application invoked in  \cite{wahba1965} is satellite attitude determination. The derivation of more sophisticated attitude estimators has been the subject of a lot of research over the last decade, mainly driven by the burst of mini UAVs (unmanned aerial vehicles), especially quadrotors. The reader is referred to, e.g., \cite{barrau2013intrinsicp} for  examples.

In the present paper we propose a general derivation of an intrinsic Cram\'er-Rao lower bound on Lie groups, that is similar to the one proposed by S.T. Smith on manifolds, that is, we retain terms up to the second order in the estimation error. The discrepancy between the estimation and the true parameter is naturally defined in terms of group operation (which makes it intrinsic). Thus, the bound differs from the Euclidean one because of the non-commutativity of the group operation, yielding some additional terms that are expressed thanks to the Lie bracket (or alternatively structure constants). It is interesting to note our result coincides with the result of S.T.  Smith in the case where the metric is bi-invariant, as in SO(3). However, both formula disagree in the general case, as the definition of estimation error in terms of group multiplication differs from the intrinsic estimation error based on the Riemannian exponential proposed by S.T. Smith. 

\section{An intrinsic Cram\'er-Rao bound on Lie groups}

We compute here the Intrinsic Cram\'er-Rao Lower Bound (ICRLB) on a Lie group $G$, up to the second order terms in the estimation error $\log(g \hat g^{-1})$, where $g \in G$ is the true value of the parameter and $\hat g\in G$ the estimate.

\subsection{Preliminaries}

Let $G$ be a Lie group of dimension $n$. To simplify notations we assume $G$ is a matrix Lie group. The tangent space  at the Identity element $Id$, denoted $\mathfrak{g}$, is called the Lie Algebra of $G$ and can be identified as $\RR^n$, that is
$$
\mathfrak{g}\approx \RR^n.
$$
The  (group) exponential map 
\begin{align*}
\exp : &\mathfrak{g}\mapsto G
\\ &\xi\to \exp(\xi),
\end{align*}provides a local diffeomorphism in a neighborhood of $Id$. The (group) logarithmic  map
$$
\log : G\mapsto \mathfrak{g},
$$
is defined as the principal inverse of exp. For any estimator $\hat g$ of a parameter $g\in G$, it allows to measure the mean quadratic estimation error projected onto the Lie algebra (the error being intrinsically defined in terms of group operation, where the group multiplication replaces the usual addition in $\RR^n$)
$$
\mathbb{E}_g \left( \log(g\hat g(X)^{-1}) \right)
$$
where $\mathbb{E}$ denotes the expectation assuming $X$ is sampled from $\mathbb{P}(X|g)$. The logarithmic map allows also to define a covariance matrix of  the estimation error:
\begin{align}
P=\mathbb{E}_g \left( \log(g\hat g(X)^{-1})\log(g\hat g(X)^{-1})^T \right)\in\RR^{n\times n}.
\label{P:def}
\end{align}

\subsection{Main result} 
Consider a family of densities parameterized by elements of  $G$
\[
p(X|g), ~X\in \RR^k,~g \in G.
\]
Using the exponential map, the intrinsic information matrix $J(g)$ can be defined in a right-invariant basis as follows: for any $\xi\in\RR^n$,
\begin{equation}
\label{eq::information}
\xi^TJ(g)\xi=\int \left( \dotex\log p \left( X\mid \exp \left( t\xi \right) g \right) \right)^T \left( \dotex\log p \left( X\mid \exp \left( t\xi \right) g \right) \right) p(X\mid g)dX,
\end{equation}
and then $J(g)$ can be recovered using the standard polarization formulas
$$
\xi^TJ(g)\nu=\frac{1}{2} \left( \left( \xi+\nu \right)^TJ(g) \left( \xi+\nu \right)-\xi^TJ(g)\xi-\nu^TJ\nu \right).
$$
Besides, using the fact that $\int p\left(X\mid \exp \left( t\xi \right)g \right)dX$ is constant (equal to $1$), which implies 
\begin{equation}
\label{eq::deriv}
0=\frac{d}{dt} \int p\left(X\mid \exp \left( t\xi \right)g \right)dX=\int \left( \frac{d}{dt} \log p \left( X\mid \exp \left( t\xi \right)g \right) \right) p \left(X\mid \exp \left( t\xi \right)g \right)dX,
\end{equation}
we have, differentiating equality \eqref{eq::deriv} a second time w.r.t $t$ and reusing that $\frac{d}{dt} p=p \frac{d}{dt} \log p$:
\begin{align*}
0= & \int \left(\frac{d^2}{dt^2}\log p \left( X\mid \exp\left(t\xi\right)g\right) \right)p\left(X\mid g \right)dX \\
 & +\int\left(\dotex\log p\left(X\mid \exp\left(t\xi\right)g\right)\right)\left(\dotex\log p\left(X\mid \exp\left(t\xi\right)g\right)\right) p(X|g) dX,
\end{align*}
allowing to recover an intrinsic version of the classical result according to which the information matrix can be also defined using a second order derivative
$$
\boxed{\xi^TJ(g)\xi=- \mathbb{E}_g \left(\frac{d^2}{dt^2}\log p\left(X\mid \exp\left(t\xi\right)g\right)\right)}.
$$
Let $\hat{g}$ be an unbiased estimator of $g$ in the sense of the intrinsic (right invariant) error $g \hat g^{-1}$, that is,
\[
\int_X \log \left(g \hat g \left(X\right)^{-1} \right) p\left(X|g\right) dX =  0.
\]
Let $P$ be the covariance matrix of the estimation error as defined in \eqref{P:def}. Our main result of this section is as follows
\begin{equation}
\label{result2}\boxed{
P \succeq  \left( Id+\frac{1}{12}P.H \right)J(g)^{-1} \left(Id+\frac{1}{12}P.H \right)^T},
\end{equation}
where we have neglected terms of order $\mathbb{E}_g \left( \norm{\log \left( g \hat g \left(X\right)^{-1} \right)}^3\right)$, 
and where $H$
is the (1,3)-structure tensor defined by 
$$H\left(X,Y,Z\right):= \left[X,\left[Y,Z\right]\right],$$and where $P.H$ is the tensor contraction of $P$ and $H$ on the two first lower indices of $H$, defined by $(P.H)_{kl}= \sum_{ij}P^{ij}H_{ijk}^l$. Using the structure constants of $G$ defined by
\begin{equation}
\label{eq::const_struct}
[e_i,e_j]:\sum_k c_{ij}^k e_k,\quad (ad_{e_i})_ k^j=c_{ij}^k,
\end{equation}
note that the components of $H$ can be expressed by the equality
\begin{equation}
\label{eq::H}
 H^{m}_{ijk}=\sum_l c_{i l}^m c_{j k}^l.
\end{equation}
The latter result is totally intrinsic, that is, it is independent of the choice of the metric in the Lie algebra $\mathfrak g$.

For small errors, we can neglect the terms in $P$ on the right hand side (curvature terms) yielding the approximation which reminds the Euclidean case
$$\boxed{P={\int_X \log\left(g\hat g\left(X\right)^{-1}\right) \log\left(g\hat g\left(X\right)^{-1}\right)^T p\left(X|g\right) dX\succeq J\left(g\right)^{-1}~\text{+ curvature terms }}}.
$$

\subsection{Proof of the result}
Let $\hat{g}$ be an unbiased estimator of $g$ in the sense of the intrinsic (right invariant) error $g \hat g^{-1}$, that is,
\[
\int_X \log \left(g \hat g \left(X\right)^{-1} \right) p\left(X|g\right) dX =  0.
\]
If we let $\xi$ be any vector of the Lie algebra and $t\in\RR$, the latter formula holds with $g$ replaced by $\exp\left(t\xi\right)g$ and $X$ sampled from $p(X|\exp\left(t\xi\right)g)$. Thus we have $\mathbb{E}_{\exp\left(t\xi\right)g}\left(\log\left[\exp\left(t\xi\right)g\hat g\left(X\right)^{-1} \right]\right) =  0$ for any $t\in \RR$. Differentiating this equality written as an integral over $X$ we get
\[
\frac{d}{dt} \int_X \log \left[ \exp\left(t\xi\right)g \hat{g}\left(X\right)^{-1}  \right] p\left(X|\exp\left(t\xi\right)g\right) dX =  0.
\]
Formally, this implies at $t=0$
\begin{equation}
\label{eq1}
\int_X \biggl( D\log|_{g \hat{g} \left( X \right)^{-1}} \left[\xi g \hat{g}(X)^{-1} \right] p\left(X|g\right) + \log \left(g \hat{g}(X)^{-1} \right) Dp\left(X\mid \cdot \right)|_g \left[\xi g\right] \biggr) dX = 0.
\end{equation}
For any linear form $u(\cdot)$ of the Lie algebra $  \frak{g}$ we have thus:
\begin{align}
-\int_X  & u \biggl( D\log|_{g \hat{g}\left(X\right)^{-1}} \left[\xi g \hat{g}\left(X\right)^{-1}  \right] \biggr) p\left(X|g\right) dX \nonumber \\
 & = \int_X u\biggl( \log \left(g \hat{g}(X)^{-1} \right)  Dp(X\mid \cdot)|_g [\xi g] \biggr) dX \nonumber \\
 & \leqslant \sqrt{ \left( \int_X u\bigl(\log \left(g \hat{g}(X)^{-1}\right) \bigr)^2 p(X|g) dX \right) \left( \int_X \bigl(D\log p(X\mid \cdot)|_g [\xi g] \bigr)^2 p(X|g) \right) dX }, \label{eq::CC}
\end{align}
where we used the Cauchy Schwarz inequality and the relationships
$$
Dp=pD\log p,\quad \text{and then} \quad p=(\sqrt p)^2.
$$
We then introduce a basis of $\frak{g}$ and the vector $\tilde{A}(X)=\log(g \hat{g}(X)^{-1})$. According to \eqref{eq::information} the right-hand integral in \eqref{eq::CC} is $J(g)$, which yields ($u$ being assimilated to a vector of $\mathfrak{g}$) :
\begin{equation}
\label{Cauchy-Schwarz}
\begin{aligned}
\biggl(u^T \int_X  &D\log|_{g \hat{g}(X)^{-1}} [\xi_{\times} g \hat{g}(X)^{-1}  ] p(X|g)dX \biggr)^2 \\
 & \leqslant \left( u^T \left[ \int_X \tilde{A}(X) \tilde{A}(X)^T p(X|g) dX \right] u \right) \biggl(\xi^T J(g) \xi \biggr), 
\end{aligned}
\end{equation}
where we added the subscript $\times$ to $\xi$, distinguishing the element $\xi_\times$ of $\mathfrak{g}$ and the column vector $\xi$ containing its coordinates in the chosen basis. Now we compute a second-order expansion  (in the estimation error) of the left-hand term. To do so, we note from the Baker-Campbell-Hausdorff (BCH) formula retaining only terms proportional to $t$
\begin{align*}
\log \left[ \exp\left(t\xi\right)Q  \right] & =\log \left[ \exp\left(t\xi\right)\exp\left(\log(Q)\right)  \right] \\
 & = t\xi - \frac{1}{2} \left[{\log(Q)},t\xi \right]+ \frac{1}{12}\left[{\log(Q)},\left[ \log(Q),  t\xi \right]+ O\left(||\log(Q)||^3\right)\right] t\xi.
\end{align*}
This gives the formula below: the second-order expansion in $\log (Q)$ of the derivative of the left-hand term w.r.t to $t$. Note that this approximation will be integrated over $Q$ in equation \eqref{eq::approx} and therefore a rigorous reasoning should prove the density $p$ is small where the higher-order terms become larger. Here we assume $p$ is sufficiently peaked for this approximation to be valid.
\begin{align}
D\log_Q [(\xi)_{\times} Q ] = [ I - \frac{1}{2} ad_{\log(Q)} + \frac{1}{12} ad_{\log(Q)}^2  ] \xi.
\end{align}
Moreover we have by linearity ($\xi$ is here deterministic):
$$
\mathbb{E}_g\left( \frac{1}{2} \left[{\log\left(g \hat{g}(X)^{-1}\right)},t\xi\right] \right)=\frac{1}{2} \left[ \mathbb{E}_g \left( \log \left(g \hat{g}(X)^{-1}\right) \right),t\xi \right]=0,
$$
and also
$$ \mathbb{E}_g [x,[x,\xi]]=\sum H_{ijk}^l \mathbb{E}_g(x_ix_j)\xi_ke_l=G^0\xi,
$$where $G^0$ is a matrix whose entries are functions of $\mathbb{E}_g(xx^T)$ and 
the structure constants (see \eqref{eq::const_struct}): $G^0$ is defined by $G^0=P.H$, i.e. $G^0_{k,l}= \sum_{ij}P^{ij}H_{ijk}^l$ with $H$ defined as in \eqref{eq::H}. We introduce the latter second-order expansion in the error in equation \eqref{Cauchy-Schwarz}:
\begin{equation}
\label{eq::approx}
\begin{aligned}
 \left[ u^T \int_X  \left[ I + \frac{1}{12} ad_{\tilde{A}(X)}^2  \right]  p(X|g) dX \xi \right]^2  \leqslant \left(u^T \left[\int_X \tilde{A}(X) \tilde{A}(X)^T p(X|g) dX \right]u \right) \biggl( \xi^T J(g) \xi \biggr).
\end{aligned}
\end{equation}
Letting $P=\int_X \tilde{A}(X) \tilde{A}(X)^T p(X|g) dX$ we get:
$$
\left[ u^T \left( I + \frac{1}{12} G^0(P)  \right) \xi \right]^2\leqslant \left(u^T P u\right) \left(\xi^T J(g) \xi\right).
$$
Replacing $\xi$ with the variable $\xi=J(g)^{-1}\left( I + \frac{1}{12} G^0(P)  \right)u$ directly allows to prove that 
\begin{equation}
\label{result}
P \succeq  \left( 1 + \frac{1}{12} G^0(P)\right) J(g)^{-1}  \left( 1 + \frac{1}{12} G^0(P)\right)^T,
\end{equation}
where
\begin{equation}
\label{eq::G(P)}
G^0(P)=\mathbb{E}_g \left[\log\left(g \hat{g}^{-1}\right),\left[\log\left(g \hat{g}^{-1}\right),\cdot \right] \right].
\end{equation}

\begin{rem}
If the model is equivariant, i.e., verifies $\forall h \in G, p\left(hX|gh^{-1}\right)=p \left(X|g\right)$ (see \cite{berger1985statistical,barrau2013note}), the study can be restricted to equivariant estimators (estimators verifying $\hat g(hX)=\hat g(X)h^{-1}$). In this case equation \eqref{eq::G(P)} simplifies:
\begin{align*}
G^0(P) & = \int_X \left[ \log\left(g \hat{g}(X)^{-1}\right),\left[\log\left(g \hat{g}(X)^{-1}\right),\cdot \right] \right] p(X|g) dX \\
 & = \int_X \left[ \log\left(\hat{g}(gX)^{-1}\right),\left[\log\left(\hat{g}(gX)^{-1}\right),\cdot \right] \right] p(gX|Id) dX \\
  & = \int_{X'} \left[ \log\left(\hat{g}(X')^{-1}\right),\left[\log\left(\hat{g}(X')^{-1}\right),\cdot \right] \right] p(X'|Id) dX',
\end{align*}
where the integration variable $X$ has been replaced by $X'=gX$ in the latter equality. Thus if the model is equivariant,   the Cram\'er-Rao bound is constant over the Lie group. 
\end{rem}

 \section{Links with the more general Riemannian manifolds case}
In the paper \cite{smith-2005}, the author derives  the following intrinsic Cram\'er-Rao bound (see Corollary 2). Assume $\theta$ lives in a Riemannian manifold and $\hat \theta$ is an unbiased estimator, i.e. $\mathbb{E}_g(\exp_\theta^{-1}\hat \theta)=0$ where $\exp$ is the geodesic exponential map at point $\theta$ associated with the chosen metric. The proposed ICRLB writes  (up to higher order terms)
$$P:=\mathbb{E}_g \left(\exp_\theta^{-1}\hat \theta \right) \left(\exp_\theta^{-1}\hat \theta \right)^T\succeq J(\theta)^{-1}-\frac{1}{3}\left(R_m(P)J(\theta)^{-1}+J(\theta)^{-1}R_m(P)\right),$$
where for sufficiently small covariance norm the matrix $R_m(P)$ can be approximated by the quadratic form
$$
\Omega \rightarrow \left<R_m(P)\Omega,\Omega \right>=\mathbb{E}_g \left< R \left(\exp_\theta^{-1}\hat \theta,\Omega \right)\Omega,\exp_\theta^{-1}\hat \theta \right> ,
$$where $R$ is the Riemannian curvature tensor at $\theta$.


The considered error $\log(g \hat{g}^{-1})$ being right-invariant, we can assume $g=Id$ to compare our result to the latter. We then see, that up to third order terms, formula \eqref{result} coincides indeed with the result of \cite{smith-2005} \textbf{if}
 $$
G^0(P)=-4R_m(P),$$where
$$ \left< R_m(P)\xi,\xi \right> = \mathbb{E}_g \left< R \left(\log \left(g \hat{g}(X)^{-1}\right),\xi \right)\xi,\log \left(g \hat{g}(X)^{-1}\right)\right>,$$
$$~\text{and}~\left< G^0(P)\xi,\xi\right>=\mathbb{E}_g \left< \left[\log\left(g \hat{g}(X)^{-1}\right),\left[\log\left(g \hat{g}(X)^{-1}\right),\xi \right] \right],\xi \right>.
 $$

If the metric is bi-invariant, as is the case for  $G=SO(3)$, we recover the result of \cite{smith-2005}. Indeed, for bi-invariant metrics on  Lie groups, the Riemannian curvature tensor satisfies for right-invariant vector fields (see ,e.g., \cite{arnold-fourier}) 
\begin{align}
 R(X,Y)Z=-\frac{1}{4}[[X,Y],Z].\label{michel}
\end{align}
The question of comparing both formulas boils down to proving for $Z$ random vector s.t. $E(Z)=0$ that $\mathbb{E}_g\langle\xi,[Z,[Z,\xi]]\rangle=-4\mathbb{E}_g\langle R(Z,\xi)\xi,Z\rangle$. It can be verified as follows:
$$
 -4\mathbb{E}_g\langle R(Z,\xi)\xi,Z\rangle=\mathbb{E}_g\langle[[Z, \xi],\xi],Z\rangle=\mathbb{E}_g\langle[Z,[Z, \xi]],\xi\rangle=\mathbb{E}_g\langle\xi,[Z,[Z,\xi]]\rangle
 $$
where the second equality stems from a property of the bi-invariant case (e.g., \cite{arnold-fourier}) that generalizes the mixed product on $SO(3)$ property, namely $\langle[X,Y],Z\rangle=\langle[Z,X],Y\rangle$. 
 
 \subsection{Differences}
 
 If the metric is not bi-invariant the results are different. This is merely because then the Lie group exponential  map and the Riemannian exponential map are not the same. Thus, our definition of the estimation error differs, so it is logical that the results be different.

\section{Conclusion}
In this paper, we have proposed an intrinsic lower bound for estimation of a parameter that lives in a Lie group. The main difference with the Euclidean case is that the estimation error between the estimate and the true parameter underlying the data is measured in terms of group multiplication (that is, the error is an element of the group). This is a much more natural way to measure estimation errors, and it has been used in countless works on statistical estimation and filtering on manifolds. But it comes at a price of additional terms in the bound, that are not easy to interpret. The bound is not closed-form, but by retaining only first and second order terms in the estimation error, we ended up with a closed form bound, that is the covariance of any unbiased estimator is lower bounded by the inverse of the intrinsic Fisher information matrix, as in the Euclidean case, plus additional terms that are functions of the covariance. Moreover, the nice structure of Lie groups allows straightforward calculations, and the bound is expressed with the help of the Lie bracket, this, in turn, being related to the sectional curvature at the identity, helping to draw a link with the general result of S. T. Smith on Riemannian manifolds \cite{smith-2005}. 

This note generalizes previous calculations \cite{barrau2015cramer} obtained on  SO(3), in the context of attitude filtering for a dynamical rigid body in space, the latter Cram\'er-Rao bound being compared to the covariance yielded by the intrinsic Kalman filter of \cite{barrau2013intrinsicp}. It would be interesting to apply the obtained general bound to pose averaging (that is on SE(3)), as ,e.g., proposed in \cite{tron2008distributed}, which could be then attacked by means of intrinsic stochastic approximation as in ,e.g., \cite{bonnabel2013stochastic} or \cite{jabu}.

Another future route could be to derive an intrinsic Cram\'er-Rao bound on homogeneous spaces. It is interesting to note that both examples of \cite{smith-2005} are homogeneous spaces. Moreover, the results on manifolds are bound to be local, but we can hope for results on Lie groups and homogeneous spaces with a large domain of validity (if not global). 
 
\subsection*{Acknowledgments}We thank Y. Ollivier for his help on some Riemannian geometry matters,  N. Boumal and P.A. Absil for the time they took to discuss with us about the subject of intrinsic Cram\'er-Rao bounds on manifolds, and J. Jakubowicz for kindly inviting us to submit this paper.


\begin{thebibliography}{10}

\bibitem{arnold-fourier}
V.I. Arnol'd.
\newblock Sur la g\'{e}om\'{e}trie diff\'{e}rentielle des groupes de {L}ie de
  dimension infinie et ses applications \`{a} l'hydrodynamique des fluides
  parfaits.
\newblock {\em Ann. Inst. Fourier}, 16:319--361, 1966.

\bibitem{barrau2013note}
Axel Barrau and Silv\`ere Bonnabel.
\newblock A note on the intrinsic {C}ram\'er-{R}ao bound.
\newblock In {\em Geometric Science of Information}, pages 377--386. Springer,
  2013.

\bibitem{barrau2015cramer}
Axel Barrau and Silv{\`e}re Bonnabel.
\newblock An intrinsic {C}ram\'er-{R}ao bound on {SO}(3) for (dynamic) attitude
  filtering.
\newblock {\em arXiv preprint, submitted}, 2015.

\bibitem{barrau2013intrinsicp}
Axel Barrau and Silv\`ere Bonnabel.
\newblock Intrinsic filtering on {L}ie groups with applications to attitude
  estimation.
\newblock {\em IEEE Transactions on Automatic Control}, 60(2):436 -- 449, 2015.

\bibitem{jabu}
A.~Bellachehab and J.~Jakubowicz.
\newblock Random pairwise gossip on {CAT}(0) metric spaces.
\newblock In {\em IEEE conference on decision and control}, pages 5593 -- 5598,
  2014.

\bibitem{berger1985statistical}
James~O Berger.
\newblock {\em Statistical decision theory and Bayesian analysis}.
\newblock Springer, 1985.

\bibitem{bonnabel2013stochastic}
Silv\`ere Bonnabel.
\newblock Stochastic gradient descent on {R}iemannian manifolds.
\newblock {\em Automatic Control, IEEE Transactions on}, 58(9):2217--2229,
  2013.

\bibitem{boumal2013intrinsic}
Nicolas Boumal.
\newblock On intrinsic {C}ram{\'e}r-{R}ao bounds for {R}iemannian submanifolds
  and quotient manifolds.
\newblock {\em IEEE transactions on signal processing}, 61(5-8):1809--1821,
  2013.

\bibitem{boumal2014cramer}
Nicolas Boumal, Amit Singer, P-A Absil, and Vincent~D Blondel.
\newblock Cram{\'e}r-{R}ao bounds for synchronization of rotations.
\newblock {\em Information and Inference}, 3(1):1--39, 2014.

\bibitem{pitman1939estimation}
EJG Pitman.
\newblock The estimation of the location and scale parameters of a continuous
  population of any given form.
\newblock {\em Biometrika}, pages 391--421, 1939.

\bibitem{smith-2005}
S.T. Smith.
\newblock Covariance, subspace, and intrinsic {C}ram\'er-{R}ao bounds.
\newblock {\em IEEE-Transactions on Signal Processing}, 53(5):1610--1629, 2005.

\bibitem{tron2008distributed}
Roberto Tron, Ren{\'e} Vidal, and Andreas Terzis.
\newblock Distributed pose averaging in camera networks via consensus on
  {SE}(3).
\newblock In {\em Distributed Smart Cameras, 2008. ICDSC 2008. Second ACM/IEEE
  International Conference on}, pages 1--10. IEEE, 2008.

\bibitem{wahba1965}
Grace Wahba.
\newblock A least squares estimate of satellite attitude.
\newblock {\em SIAM review}, 7(3):409--409, 1965.

\end{thebibliography}
\end{document}